\documentclass[pre,aps,draft,showpacs]{revtex4}

\usepackage[dvips,final]{graphicx}
\usepackage{bm}

\begin{document}

\title{The electroclinic effect and modulated phases in smectic liquid crystals}
\author{Robert B. Meyer}
\affiliation{The Martin Fisher School of Physics, Brandeis University, Waltham, Massachusetts 02254-9110}
\author{ Robert A. Pelcovits}
\affiliation{Department of Physics, Brown University, Providence, RI 02912}
\date{\today}

\begin{abstract}

We explore the possibility that the unusually large electroclinic
effect observed in the smectic--{\textit A} phase of a
ferroelectric liquid crystal arises from the presence of an
ordered array of disclination lines and walls in a
smectic--$C^\ast$ phase. If the spacing of these defects is in the
subvisible range, this modulated smectic--$C^\ast$ phase would be
similar macroscopically to a smectic--{\textit A} phase. The
application of an electric field distorts the array, producing a
large polarization, and hence a large electroclinic effect. We
show that with suitable elastic parameters and sufficiently large
chirality, the modulated phase is favored over the smectic--{\it
A} and helically twisted smectic--$C^\ast$ phases. We propose
various experimental tests of this scenario.
\end{abstract} \pacs{61.30.Cz,  61.30Gd,  61.30Mp} \maketitle
\section{\label{sec:Introduction}Introduction}

\begin{figure}
\includegraphics[scale=0.5]{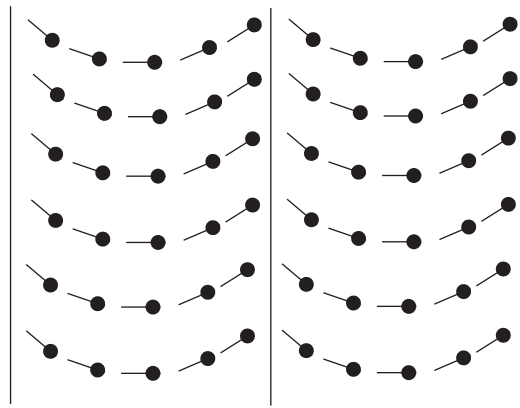}
\caption{\label{striped} Striped modulated phase in a thin
Sm--$C^\ast$ film. The circular heads of the objects shown denote
the heads of the local $\bm c$ director; the vertical lines are
domain walls where the $\bm c$ director rotates by an angle of
order $\pi$.}
\end{figure}

\begin{figure}
\includegraphics[scale=0.5]{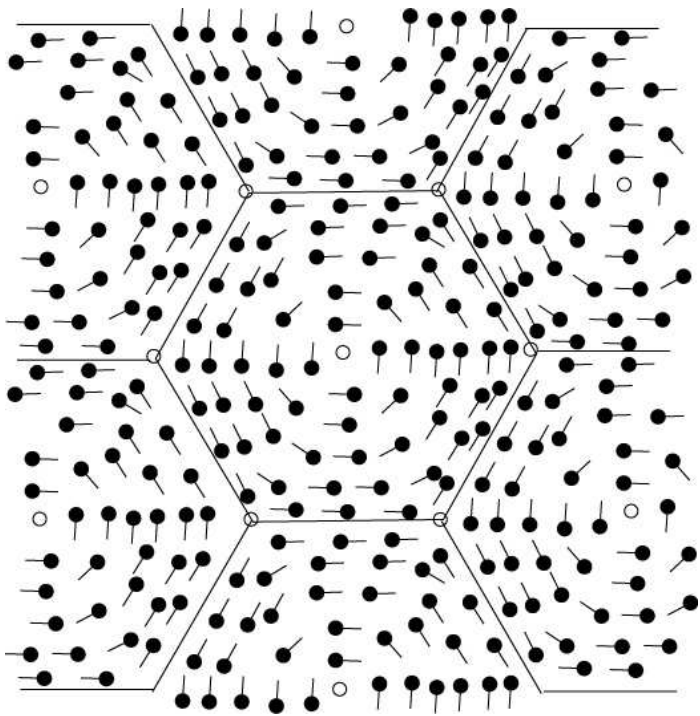}
\caption{\label{hex} Hexagonal modulated phase in a thin
Sm--$C^\ast$ film. There is a +1 disclination at the center of
each unit cell, and $-1/2$ disclinations at each corner. Domain
walls where the $\bm c$ director rotates by an angle of order
$\pi$ bound the cell.}
\end{figure}

Chiral liquid crystals exhibit a variety of interesting phenomena
\cite{Kamien}. Chiral molecules tend to pack at a slight angle
with respect to their neighbors. This preferred packing leads to
spontaneous twist in the molecular orientation, producing phases
such as the cholesteric and helically twisted smectic--$C^\ast$
(Sm--$C^\ast$). In the case of two--dimensional Sm--$C^\ast$
films, the chiral modulation of the molecular orientation cannot
occur in a defect--free fashion. Instead, for large chirality, the
system will form a modulated phase consisting of a regular network
of defect walls and possibly points. One possible modulated phase
is a ``striped'' phase consisting of a regular array of parallel
domain walls (Fig. \ref{striped}). In between these walls the
{\textit c} director rotates in the sense preferred by the
chirality. Another possible modulated phase consists of a
hexagonal array of domain walls, with point defects located at the
centers and corners of the hexagonal cells (Fig. \ref{hex}).


Chirality also leads to unusual effects in orientationally
disordered phases. As shown by Garoff and Meyer \cite{Garoff}
using symmetry arguments, if an electric field is applied to a
system of chiral molecules in the Sm--{\textit A} phase, an
average molecular tilt will be produced, a response termed the
electroclinic effect. These symmetry arguments do not, however,
indicate the microscopic basis of the electroclinic effect. To
date, two microscopic models have been proposed. In the first, the
long axes of the molecules are assumed to be perpendicular to the
smectic layers in the Sm--{\textit A} phase. The electric field
causes the molecules to tilt, with an accompanying significant
change in the smectic layer spacing. In the second scenario
suggested by de Vries \cite{deVries}, the molecules in the
Sm--{\textit A} phase are assumed to be tilted even in the absence
of the electric field, but azimuthally disordered. The electric
field then induces azimuthal order, with no significant change in
the smectic layer spacing. Experimental studies have shown
evidence for both models, at least in terms of the field--induced
layer contraction. Recently Selinger et al.~\cite{Selinger} noted
that the optical birefringence of the liquid crystal will also
distinguish between the two microscopic pictures. Because the
field induces orientational order in the second model while simply
tilting the already aligned molecules in the first, the
birefringence should change dramatically in the former model, but
not the latter. It appears that each of these models has
experimental validity, namely there are materials with substantial
layer contraction and weak field dependence in the birefringence,
and other materials with the opposite properties. However, even a
detailed theory based on the second model \cite{Selinger} is
unable to fully describe the experimental data. Furthermore, many
materials show an unusually large electroclinic effect, well into
the Sm--{\textit A} phase. While it is tempting to associate this
large effect with strong critical fluctuations \cite{Selinger}, a
completely different molecular mechanism may be relevant.

In this paper we propose a new model for the microscopic origin of
the electroclinic effect, one that shares some of the macroscopic
characteristics of the de Vries scenario, specifically, no layer
contraction and field dependent birefringence. We postulate that
what is assumed to be a Sm--{\textit A} phase in some materials
close to the Sm--$C^\ast-A$ transition, could in fact be a
modulated version of the smectic--$C^\ast$ phase, a network of
defects, walls and lines, but one where the spacing between the
defects is in the subvisible range. This modulated phase would
have no long--range azimuthal order or net polarization in the
absence of an electric field. Thus, ``to the eye'', this phase
would appear to be a Sm--{\textit A} phase. However, the
application of an electric field distorts the structure of the
defect array and leads to a large electroclinic effect. We now
describe our proposed modulated phase and its response to an
electric field in detail.

A bulk chiral Sm--$C^\ast$ can be described by the following
elastic free energy \cite{deGennes}:
\begin{eqnarray}
 F &=& \int d^{3} x \left[ \frac{1}{2}\;K_1(\bm
{\nabla \cdot c})^2 + \frac{1}{2}\;K_2(\bm{ c \cdot \nabla \times
c} + q)^2+ \frac{1}{2}\;K_{13}({\bm {\hat z \cdot \nabla \times
c}})(\bm{ c \cdot \nabla \times c}) \right. \nonumber \\ & +
&\left. \frac{1}{2}\; K_3({\bm {\hat z \cdot \nabla \times c}})^2
- w \;{\bm {\hat z \cdot \nabla \times c}}\right] \label{F}
\end{eqnarray}
where $\bm c$ is the projection of the nematic director on the
plane of the layers. We assume that the molecular tilt is uniform
and thus $\bm c$ is assumed to be of fixed length. Chirality
appears in two terms in this free energy. First, chirality induces
a spontaneous twist of the $\bm c$ director as indicated in the
second term on the right hand side of Eq.~(\ref{F}), which is
minimized in a helical state with pitch $2\pi/q$. Second,
chirality induces spontaneous bend within each smectic layer as
indicated by the last term in the free energy, where we assume $w
> 0$. The free energy as written is zero for the perfect helical
state with pitch $2 \pi/q$. Because the smectic--$C^\ast$ phase is
ferroelectric, with the spontaneous polarization perpendicular to
$\bm c$, we must add to this free energy an electrostatic term
arising from space charge density proportional to ${\bm {\hat z
\cdot \nabla \times c}}$. We ignore this term for now, since we
will find that in our proposed modulated phase, it adds only a
constant to the free energy density, independent of the detailed
structure of our model.

Considerable work has been done on analyzing Eq.~(\ref{F}) in
{\textit two}--dimensional chiral smectic liquid crystals, where
the terms proportional to $K_2$ and $K_{13}$ in the free energy
above are absent, and thus, there will be no competition between
the two chiral terms. While only one chiral term is present, it is
impossible to impose a spontaneous bend throughout the system.
Defects must be introduced which separate domains of spontaneous
bend, thus leading to a modulated phase.  Two types of modulated
phases have been proposed, a ``striped'' phase (Fig.
\ref{striped}) with a one--dimensional modulation
\cite{Langer,Petschek1,Petschek2}, and a hexagonal phase (Fig.
\ref{hex}) with a two--dimensional modulation. In the former case,
parallel domain walls separate regions of spontaneous bend, while
the latter consists of a two--dimensional array of domain walls
and point disclinations. A $+1$ point disclination lies at the
center of each hexagonal cell, while the boundaries of the cell
are domain walls where the $\bm c$ director rotates by $\pi$ (a
description of internal structures of these walls in terms of a
nonfixed length $\bm c$ director field is given in
\cite{Petschek1,Petschek2}). A $-\frac{1}{2}$ point disclination
is located at each corner of the cell. The hexagonal phase which
involves nearly pure bend is favored over the striped phase when
$K_1 \gg K_3$. The size of the unit cell is inversely proportional
to the strength of the chirality parameter $w$. The free energy of
the hexagonal lattice includes (see Eq. (\ref{f}) below) an
attractive logarithmic interaction between the $+1$ disclination
at the center of the cell and the $-\frac{1}{2}$ disclinations at
the corners. Thus, strongly chiral materials, where the unit cell
is small, will favor this interaction, and hence the hexagonal
phase \cite{Kamien}.

Returning to the full three--dimensional free energy,
Eq.~(\ref{F}), we consider the possibility of a modulated phase. A
bulk modulated phase must somehow compromise between the two
competing chirality--induced tendencies: spontaneous in--layer
bend and spontaneous twist about the layer normal. To describe the
electroclinic effect in our proposed scenario the bulk modulated
phase must have zero net polarization. Thus, a bulk modulated
phase produced by simply extending the two--dimensional striped
phase in the third direction is not acceptable, even if the pitch
of the spontaneous twist were assumed to be infinitely large. A
three--dimensional model for a thick chiral smectic film was
proposed in Ref. \cite{Gorecka}, where the in--plane modulation
and domain walls are confined to surface regions, and the director
twists in the interior of the film. Here we consider a
space--filling modulated structure, with no need for surface
structures.

Our starting point in constructing a suitable bulk modulated phase
is the director configuration of the $+1$ point disclination at
the center of the two--dimensional hexagonal cell (Fig.
\ref{hex}). Expressing the $\bm c$ director as ${\bm
c}=(\cos{\theta(x,y)},\sin{\theta(x,y)})$, the angle $\theta$ is
given by:

\begin{equation}
\label{theta}
\theta(x,y)= \phi(x,y) + \phi_0,
\end{equation}
where the polar angle $\phi(x,y)$ is given by
$\phi(x,y)=\arctan{(y/x)}$, and $\phi_0=\pi/2$ (for the sign of
the chirality appropriate to the structure shown in  Fig.
\ref{hex}; for the opposite sign of chirality, $\phi_0=-\pi/2$,
and the director circulation will be clockwise). This
configuration represents pure bend and describes the director
configuration away from the domain walls which bound the unit
hexagonal cell. We now form a three--dimensional simple hexagonal
lattice, with a unit cell of height $h$, and cross--sectional area
as in the two--dimensional case. We specify the director
configuration as follows. In the midplane of the unit cell (taken
to be the $z=0$ plane) we assume the perfect bend director
configuration of Eq. (\ref{theta}). We then introduce twist in the
$z$ direction, by writing the angle $\theta(x,y,z)$ of the $\bm c$
director as follows:

\begin{equation}
\label{thetaz}
\theta(x,y,z)= \phi(x,y) + \phi_0(z),
\end{equation}
with

\begin{equation}
\phi_0(z)= \pi/2 + \pi z/\ell, \ \ \ -h/2<z<h/2.
\end{equation}
The pitch $\ell$ of this helical structure  will be determined by
minimizing the free energy density of the unit cell.  With this
model for the modulated phase, the $K_{13}$ and $w$ terms in the
free energy, Eq. (\ref{F}), can be combined into a term like the
$w$ term with $w_{eff}=w+\pi/\ell$.  We find that $\ell$ is so
large that for simplicity we set $w_{eff} = w$.

\begin{figure}
\includegraphics[scale=0.5]{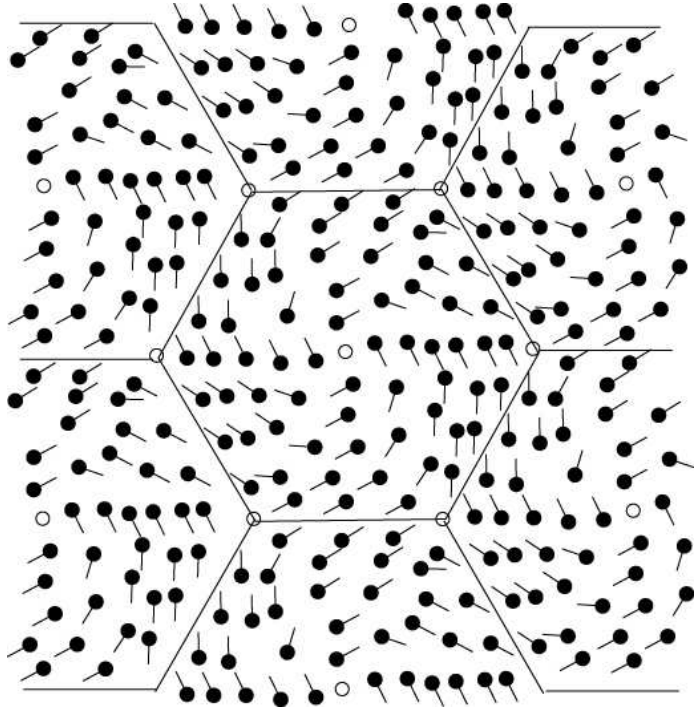}
\caption{\label{twist} A slice through the bulk hexagonal lattice
at $z=\ell/6$, showing how the director has twisted by an angle of
$\pi/6$ relative to the $z=0$ slice shown in Fig.~\ref{hex}.}
\end{figure}

Thus the unit cell has a $+1$ disclination line running along the
central  $z$ axis of the cell and parallel $-1/2$ disclination
lines along the edges of the cell. Within the cell the director
twists along the $z$ axis, in the sense preferred by the chirality
(see Fig.~\ref{twist}). The top and bottom faces of the cell
(perpendicular to the $z$ axis) are domain walls where the
director undergoes a rotation by $2\pi h/\ell$. The six other
faces of the cell are walls where the $\bm c$ director rotates by
$\pi$.

The free energy density of the hexagonal unit cell (of side $R$)
relative to the helical state is given by:

\begin{equation}
\label{f} {\textit {f}}= \frac{2\pi K\ln(R/R_o) +
K^\prime}{3\sqrt{3}R^2/2} + K_2\biggl(q-\frac{2\pi}{\ell}\biggr)^2
-\frac{2\ln{3}w\ell}{\pi hR} \sin(\pi h/2 \ell)+\frac{2A_T}{h}
+\frac{2A_S}{\sqrt{3}R}\; ,
\end{equation}
where we have assumed that $K_1=K_3=K$ to simplify the elastic
energy expression associated with the disclination lines. The core
size of the disclination lines is denoted by $R_o$, and we
represent the domain wall energy per unit area of the top and
bottom faces of the cell by $A_T$, and the wall energy per unit
area of each side face by $A_S$.  The term proportional to
$\ln(R/a)$ includes the elastic energy per unit cell associated
with the $+1$ disclination line at the center of the cell and the
surrounding $-1/2$ disclination \cite{Petschek2}. The term
$K^\prime$ represents the remainder of the splay--bend elastic
energy per unit cell. The term proportional to $w$ is obtained by
integrating  the corresponding term in Eq.~(\ref{F}) over the
hexagonal unit cell using the director profile Eq.~(\ref{thetaz}),
thus ignoring the distortion of this profile near the edges of the
cell.

\begin{figure}
\includegraphics[scale=0.5]{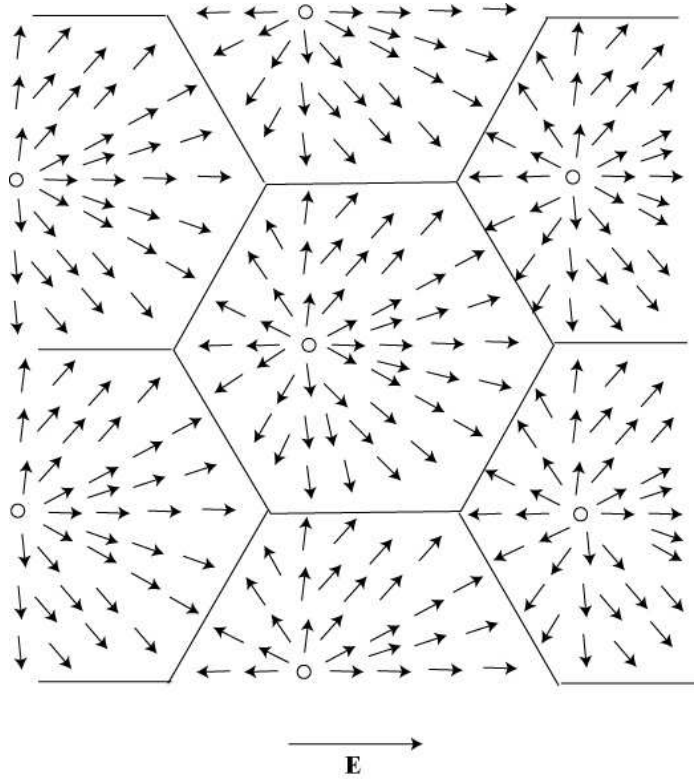}
\caption{\label{fig:P} The $z=0$ slice through the bulk hexagonal
lattice in the presence of an electric field in the $+x$
direction. The arrows here represent the molecular dipole moments
which are locally perpendicular to the $\bm c$ director and lie in
the $x-y$ plane. The electric field shifts the disclination line
at the center of each cell to induce a net polarization in the
direction of the field.}
\end{figure}

Before minimizing this free energy density we consider the
response of the modulated state to an electric field, i.e., the
electroclinic effect. We assume that the direction of the
molecular dipole moment $\bm p$ is given by $\bm{c \times \hat
z}$. In the two--dimensional hexagonal modulated phase (see Fig.
\ref{hex}), the molecular dipole moments near the center of the
cell will point radially outward, given the bend configuration of
the $\bm c$ director. Thus, an electric field applied in the $+x$
direction will shift each $+1$ disclination in the opposite
direction, inducing a net polarization in the $+x$ direction, as
shown in Fig. \ref{fig:P}.
For the three--dimensional cell we assume that the $+1$
disclination lines at the cell centers also shift rigidly in the
opposite direction to the field, ignoring any possible bending of
the line due to the twist of the director in each cell (this
should be a higher--order effect in the electric susceptibility).
Assuming that the disclination lines shift by a distance $d$ in
the $-x$ direction, we find that the polarization (the net dipole
moment of a cell divided by its volume $V$) is given to leading
order in $d$ by:

\begin{equation}
\label{P}
P= \frac{2d p \ell R}{V a^3} \sin(\pi h/2\ell),
\end{equation}
where $p$ is the molecular dipole moment and $a$ is the average
intermolecular spacing. In deriving $P$ we have assumed that the
unit cell is a cylinder of radius $R$ centered on the original
position of the line; ignoring the hexagonal shape of the cell
merely changes the numerical prefactor appearing in the above
expression.

The distance $d$ is determined by minimizing the change in the
free energy density due to the presence of the field. Displacing
the disclination line by a distance $d$ costs elastic energy
proportional to $K h (d/R)^2$. Thus, $d$ is given by:

\begin{equation}
d=\frac{p \ell E R^3}{K h a^3} \sin(\pi h/2 \ell),
\end{equation}
the polarization by:

\begin{equation}
P\approx\frac{ p^2 \ell^2 R^2 E}{K a^6 h^2} \sin^2(\pi h/2 \ell),
\end{equation}
and the susceptibility by:
\begin{equation}
\label{chiP} \chi_P\equiv\frac{dP}{dE}=\frac{ p^2 \ell^2 R^2 }{K
a^6 h^2} \sin^2(\pi h/2 \ell).
\end{equation}

We now consider the minimization of $f$, which is a function of
$R,\ell,$ and $h$. Minimizing first with respect to $R$, we find
the following condition:
\begin{equation}
\label{R}
R={4\pi K \over 3 f_W} \ln(R/R_o^\prime),
\end{equation}
where
\begin{equation}
f_W={w \ell\sqrt{3} \ln{3}\over \pi h}  \sin(\pi h/2 \ell)- A_S,
\end{equation}
and
\begin{equation}
R_o^\prime=R_o\exp\Bigl(\frac{1}{2} - \frac{K^\prime}{2 \pi K}\Bigr).
\end{equation}

Eq.(\ref{R}) has a solution for $R>R_o^\prime$ providing $f_W>0$.
This condition is physically equivalent \cite{Petschek1} to
requiring that the domain wall energy of the side walls of the
cell be negative in a non--fixed length director model, where the
wall energy includes surface contributions, i.e., contributions
from the total gradient term $\bm{\nabla \times c}$. Numerical
minimization of the free energy density, Eq. (\ref{f}), with
respect to $R, \ell,$ and $h$, (choosing $K=K_2=1,
K^\prime=A_T=A_S=0.1$) indicates that for values of $w$ that yield
$f<0$, (i.e., when the modulated state is favored over the helical
state), $h \approx 10^5,\ell/h \approx 100$, and:
\begin{equation}
\label{approx}
\frac{2 \ell}{\pi h} \sin(\pi h/2 \ell) \approx 1.
\end{equation}
The condition $\ell \gg h$ is sensible, because in that case the
director pattern in each cell will be nearly pure bend. In this
limit the condition $f_W>0$ implies that  $A_S < w \sqrt{3} \ln{3}
/2 \approx w$.

With $\ell$ very large, the polarization space charge density,
proportional to ${\bm {\hat z \cdot \nabla \times c}}$ varies in
each cell as $1/r$, with $r$ the distance from the central
disclination, and is almost independent of $z$.  This produces an
electric field ${\bm E}$ which is almost purely radial, and
constant in magnitude, considering the cell to be approximated by
a cylinder. At the outer boundary of the cell there is a surface
charge which precisely cancels the charge in the interior of the
cell, so that there is no electric field outside each cell due to
the charge distribution within the cell and on its boundaries.
Thus the electrostatic energy density can be evaluated in a single
cell, and is proportional to $E^2$, independent of the radius or
height of the cell.  Although there are also ionic charges present
in liquid crystals, typical estimates of their screening length
indicate that they would not alter the electric fields at the
length scales we are considering.

Assuming that $\ell$ is large, as indicated by the numerical
minimization of Eq. (\ref{f}), (specifically, $\ell \gg q^{-1}$),
and using Eq.~(\ref{approx}), the free energy density Eq.
(\ref{f}) can be approximated as:
\begin{equation}
\label{fapprox} f \approx K_2 q^2 +\frac{P_0^2}{8\pi\epsilon}-
{\sqrt{3} f_W^2 (2 \ln(R/R_o^\prime) -1)\over 8 \pi K
\ln^2(R/R_o^\prime)}.
\end{equation}
in which the electrostatic term, with spontaneous polarization
density $P_0$ and dielectric constant $\epsilon$, has been
included.

\begin{figure}
\includegraphics[scale=0.5]{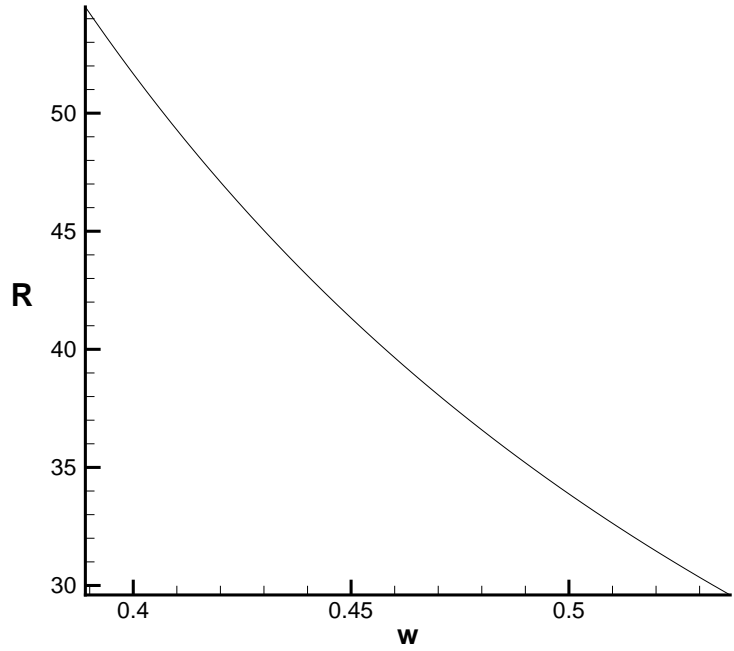}
\caption{\label{fig:R} The side $R$ of the hexagonal unit cell
determined by numerically solving Eq.~(\ref{R}), using Eq.
(\ref{approx}), and choosing $K=1,K^\prime=A_S=0.1$ and $K_2 q^2
+P_0^2/(8\pi\epsilon) = 0.0025$.}
\end{figure}

\begin{figure}
\includegraphics[scale=0.5]{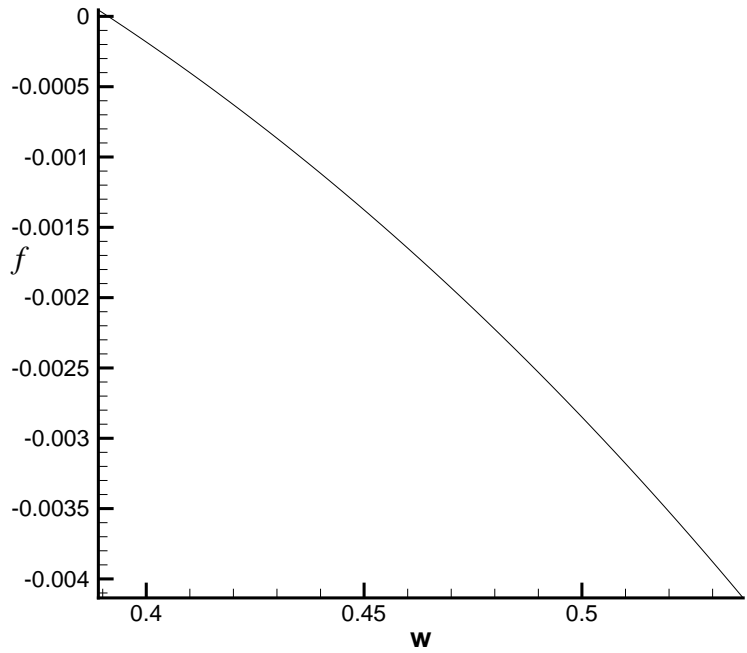}
\caption{\label{fig:f} The approximate free energy density
Eq.~(\ref{fapprox}) as a function of $w$. The side $R$ of the unit
cell is determined numerically from Eq.~(\ref{R}). The parameters
used in Eqs.~(\ref{R}) and (\ref{f}) are: $K=K_2=1,
K^\prime=A_T=A_S=0.1$ and $K_2 q^2 +P_0^2/(8\pi\epsilon) =
0.0025$.}
\end{figure}

We determine $R$ by numerically solving Eq.~(\ref{R}), using Eq.
(\ref{approx}), and choosing $K=1,K^\prime=A_S=0.1$ and $K_2 q^2
+P_0^2/(8\pi\epsilon) = 0.0025$ ; the results are shown in Fig.
\ref{fig:R}.
Using these results for $R$ we obtain the approximate free energy
density, Eq.~(\ref{fapprox}), as a function of $w$ (as shown in
Fig. \ref{fig:f}, choosing $K_2=1, A_T=0.1$). We see that if $w
\gtrsim 0.39$, the hexagonal phase is favored over the helical
state. The numerical minimization of the full expression for the
free energy, Eq.~(\ref{f}), indicates that in this phase $h
\approx 10^5$, in units of an intermolecular spacing. Thus, the
hexagonal phase in a real system would likely consist of a single
lattice plane of hexagonal cells, i.e., the system is effectively
two--dimensional.

\begin{figure}
\includegraphics[scale=0.5]{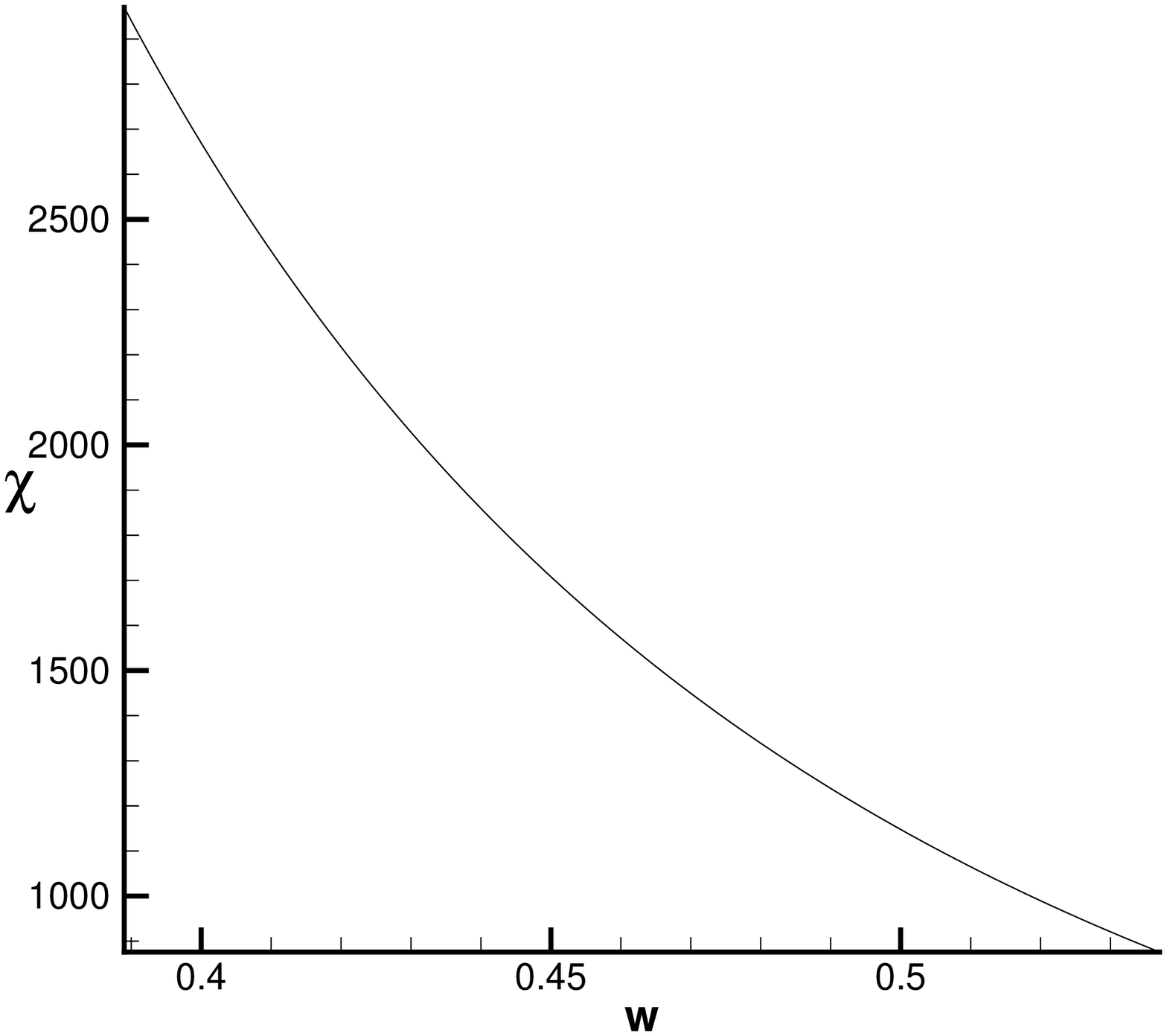}
\caption{\label{chi} The susceptibility $\chi_P$ in arbitrary
units as a function of $w$, determined from Eq.~\ref{chiapprox})
and the numerical solution of Eq.~\ref{R}).}
\end{figure}

From Eqs. (\ref{chiP}) and (\ref{approx}) we see that:
\begin{equation}
\label{chiapprox}
\chi_P \sim R^2.
\end{equation}
We plot $\chi_P$ as a function of $w$ in Fig. \ref{chi}.

\section{\label{sec:Conclusions}Conclusions}

Returning now to the observations that motivated this work, we can
see that if $R$ is a sub-visible length, then the modulated phase
would resemble a Sm--{\textit A} phase, optically.  In our model,
the apparent Sm--{\textit A} to Sm--$C^\ast$ transition, as
temperature is lowered, would actually be mimicked by the gradual
change of physical parameters, so that below the transition
temperature, the helical Sm--$C^\ast$ phase would have lower free
energy than the modulated phase.

In our model, this would be a first order transition. Since the
driving term in the free energy density, Eq.~(\ref{f}), favoring
the modulated phase is the term proportional to $w$, one could say
that if $w$ decreases as temperature is lowered while all other
parameters remain constant, then when $w$ passes through a
particular value (about 0.39 in our calculated example) a first
order transition to the helical Sm--$C^\ast$ phase would occur.
Approaching this transition, $R$ would be increasing, as would the
susceptibility, $\chi_P$, as seen in the figures above.

The value of $R$ could well remain sub-visible throughout the
range of the modulated phase.  The electroclinic effect could be
large throughout the temperature range of the modulated phase,
with a rapid increase as the transition to the helical
Sm--$C^\ast$ phase is approached. The mean smectic layer spacing
would be almost constant through the phase transition.  In the
modulated phase it would be governed by an average between the
layer spacing in the bulk and that in the side walls of the
hexagonal cells, in which it is probably larger, due to a smaller
molecular tilt angle. Therefore, at the phase transition, when the
walls disappear, a very slight decrease in mean layer spacing
would result. Qualitatively, these features agree with
observations.

One could try to orchestrate a more physically appealing model of
how the parameters of the free energy might change with
temperature.  For instance, $A_S$ steadily increasing would have
essentially the same result as $w$ steadily decreasing. More
generally, any combination of $K_2 q^2 +P_0^2/(8\pi\epsilon)$
increasing and/or $f_W$ decreasing with falling temperature could
produce the qualitative features of the apparent Sm--{\textit A}
phase and the phase transition.

The most valuable test for the validity of this model in real
materials would be the direct observation of the periodic
structure of the modulated phase.  The first appropriate
sub-optical probe that comes to mind is small angle X-ray
diffraction, especially resonant X-ray diffraction that is most
sensitive to variations of molecular orientation in a structure
\cite{resonant}. Hiroshi Yokoyama \cite{HY} has pointed out to us
that atomic force microscopy might also be a suitable probe, if
the modulated phase structure is preserved in a thin layer
deposited on a substrate.

Remarkably, in a recent paper \cite{bookshelf} on one of the
materials whose properties stimulated this work, optical
observations consistent with a modulated phase, just at the
transition from the Sm--{\textit A} to the Sm--$C^\ast$ phase, are
reported.  In a thin layer of material in the ``bookshelf''
geometry, i.e., the smectic layers viewed edge-on, thin parallel
stripes are observed running perpendicular to the smectic layers.
These stripes are possibly consistent with the long thin hexagonal
cells that we propose here.

\section*{Acknowledgments}

We thank  R. Pindak for helpful discussions.
This work was supported by the National Science Foundation under grants
DMR--9873849 and DMR--9974388.

\end{document}